\documentclass[prd,twocolumn,nofootinbib]{revtex4}

\usepackage{graphicx}
\usepackage{dcolumn}
\usepackage{bm}
\usepackage{color}
\usepackage{hyperref}
\usepackage{amsmath, mathtools}
\usepackage{amssymb}
\usepackage{wasysym}
\usepackage{graphicx} 
\usepackage{booktabs} 
\usepackage{pdfsync}
\usepackage{verbatim}
\usepackage{latexsym}
\usepackage{dsfont}

\def\sec#1{\section{#1} }
\def\ssec#1{\subsection{#1} }

\def\({\left(}
\def\){\right)}
\def\[{\left[}
\def\]{\right]}

\def\f#1#2{\frac{#1}{#2}}
\def\g{\gamma}

\def\L{\Lambda}

\def\p{\pi}
\def\r{\rho}

\def\s{\sigma}

\def\th{\theta}
\def\x{\xi}

\def\cmsg{\text{~cm}^2/\text{g}}
\def\gcmc{\text{g/cm}^3}

\def\GeV{\text{GeV}}

\def\cm{\text{cm}}

\def\<{\langle}
\def\>{\rangle}

\providecommand{\abs}[1]{\left\lvert#1\right\rvert}

\def\sX{\s_\text{\tiny X}}

\def\MX{M_\text{\tiny X}}

\def\nX{n_\text{\tiny X}}
\def\rX{\r_\text{\tiny X}}
\def\vX{v_\text{\tiny X}}

\def\sm{\sX/\MX}

\begin{document}

\author{David M. Jacobs}
\email{dm.jacobs@uct.ac.za}
\affiliation{Astrophysics, Cosmology and Gravity Centre,\\
Department of Mathematics and Applied Mathematics,\\
University of Cape Town\\
Rondebosch 7701, Cape Town, South Africa}

\author{Glenn D. Starkman}
\email{glenn.starkman@case.edu}
\affiliation{CERCA, Physics Department, Case Western Reserve University\\Cleveland, Ohio 44106-7079, USA}%

\author{Amanda Weltman}
\email{amanda.weltman@uct.ac.za}
\affiliation{Astrophysics, Cosmology and Gravity Centre,\\
Department of Mathematics and Applied Mathematics,\\
University of Cape Town\\
Rondebosch 7701, Cape Town, South Africa}

\title{Resonant bar detector constraints on macro dark matter}

\begin{abstract}
The current standard model of cosmology, $\L$CDM, requires dark matter to make up around $25\%$ of the total energy budget of the Universe. Yet, quite puzzlingly, there appears to be no candidate particle in the current Standard Model of particle physics.  Assuming the validity of the cold dark matter (CDM) paradigm, dark matter has evaded detection thus far either because it is intrinsically a weakly interacting substance or because its interactions are suppressed by its high constituent mass and low number density. Most approaches to explain dark matter to date assume the former and therefore require beyond-the-Standard-Model particles that have yet to be observed directly or indirectly. Given the dearth of evidence for this class of candidates it is timely to consider the latter possibility, which allows for candidates that may or may not arise from the Standard Model. In this work we extend a recent study of this general class of so-called \emph{macro dark matter}--candidates with characteristic masses of grams and geometric cross sections of $\cm^2$. We consider new bounds that can be set using existing data from the resonant bar gravitational wave detectors NAUTILUS and EXPLORER. 
\end{abstract}


\maketitle
\flushbottom

\sec{Introduction}

The last two decades have ushered in a new era of precision cosmology with a plethora of modern experiments and observations all leading to the so-called $\L$CDM concordance model of the Universe. A crucial component of this model is the cold dark matter (CDM) making up around 25\% of the total energy density of our Universe, as inferred from observations on large scales (see e.g. \cite{Ade:2013zuv, Anderson:2013zyy, Suzuki:2011hu, Conley:2011ku}). On intermediate scales its existence is observed indirectly due to the discrepancies between the gravitational and luminous masses of large astrophysical systems. The dark matter cannot be made of ordinary baryonic matter (see e.g. \cite{Olive:1980bu, Yang:1983gn, Freese:1999ge}, and references therein) nor can it consist of any fundamental particle of the Standard Model of particle physics.

Although a tremendous effort has been devoted to the study of particle candidates that are intrinsically weakly interacting, such as weakly interacting massive particles and axions, alternative dark matter candidates deserve increased attention given the lack of direct detection of those canonical possibilities to date (see e.g. \cite{Akerib:2013tjd, Arik:2013nya}).  Other candidates can couple to Standard Model particles with high probability but are nevertheless \emph{effectively} weakly interacting or ``dark" because they are very massive, and therefore have a lower number density.  Several examples exist in the literature, most notably nuclear-dense candidates with a Standard Model basis (e.g. \cite{Witten:1984rs, Lynn:1989xb, Lynn:2010uh}).   Other similarly massive candidates include compact objects with some connection to the Standard Model (e.g. \cite{Zhitnitsky:2002nr, Zhitnitsky:2006vt, Labun:2011wn}), primordial black holes \cite{Carr:1974nx}, and other candidates that may be found in the literature (e.g. \cite{Kusenko:1997si, Khlopov:2013ava, Murayama:2009nj, Derevianko:2013oaa, Stadnik:2014cea}).

A wide range of Earth-based constraints on nuclear-dense candidates (nuclearites) was considered in \cite{DeRujula:1984ig} and more recent bounds were presented in \cite{Burdin:2014xma}.  Such constraints are only made for candidates with a fixed internal mass density of roughly $3.5\times10^{14}~\gcmc$, and therefore they obey the specific cross section--mass relation $\sX\propto\MX^{2/3}$. Recently, however, a model-independent study of the general class of ``macroscopic"  candidates has been considered in \cite{Jacobs:2014yca}. These macro dark matter candidates, referred to as \emph{Macros}, have characteristic masses and geometric cross sections of grams and $\cm^2$ and \emph{may or may not have a Standard Model origin}. A number of Earth-based, astrophysical, and cosmological observations were used in \cite{Jacobs:2014yca} to place constraints on Macros.

Here we improve upon the constraints found in \cite{Jacobs:2014yca} by extending existing work on nuclearite constraints from resonant bar gravitational wave detectors for application to macro dark matter. The use of large aluminum bars as detectors of gravitational waves has been studied for more than five decades \cite{Weber:1960zz}, and their sensitivity to cosmic rays has been appreciated for some time \cite{Ezrow:1970yg}. Because of the thermoacoustic effect and their resonant properties, such aluminum bars are also sensitive (when cryogenically cooled) to other exotic cosmic impactors, such as monopoles \cite{Bernard:1983wj} and nuclearites \cite{Liu:1988wn}. The resonant bar experiments NAUTILUS \cite{Astone:1997gi}, which ran from 2003--2012 in Frascati, Italy, and EXPLORER \cite{Astone:1992zf}, which ran from 2003--2009 at CERN, were recently used to constrain nuclearite dark matter with masses up to $\sim 10^{-4}$ g \cite{Astone:2013xna}.

Compared to nuclearites, macro dark matter has the added complexity that both the cross section and mass are independent parameters, and both of these determine Macro detectability. Macros of a fixed mass with larger cross sections, for example, will experience more drag and so would arrive at a detector with less energy; they would therefore have a lower probability to reach a detection threshold. Here we expand on the work of \cite{Bernard:1983wj} and  \cite{Liu:1988wn}, to compute the detectability of macro dark matter as a function of both its cross section and mass. We then use the null results of \cite{Astone:2013xna} to constrain a portion of the Macro parameter space and improve on the constraints found in \cite{Jacobs:2014yca} on the elastic scattering cross section of macro dark matter with baryons.


\sec{Macro Detection via Resonant Bar Detectors}

\ssec{Macro velocity evolution}

Given an incoming Macro galactic velocity, $v_0$, its evolution as it impacts the Earth as a function of the reduced cross section, $\sm$, is approximately given by
\begin{equation}\label{Macro_velocity}
\vX(r)=v_0\, e^{-\f{\sX}{\MX}\<\r r\> }\,,
\end{equation}
where $\<\r r\>$ is the column density encountered by the Macro.  
This column density can have three distinct contributions, the first being atmospheric.
 Simple atmospheric models give an atmospheric density that depends exponentially on altitude:
\begin{equation}
\r(z)=\r_0 e^{-z/H}\,,
\end{equation}
where $z$ is the altitude above sea level, $\r_0\simeq10^{-3}\, \gcmc$, and the scale height $H\simeq10$ km. Writing $z=r \cos{\th}$, where $\th$ is the impact angle, the column density encountered through the atmosphere is
\begin{align}
\<\r r\>_\text{\tiny atm}&= \int_{r}^{\infty} dr'\, \r\notag\\
&\simeq\f{\r_0 H}{\cos{\th}} e^{-z/H}\,.
\end{align}
In addition, we shall conservatively assume that at angles $\th>\p/3$ the Macro would encounter an additional column density of 100 meters water equivalent (MWE), or
\begin{align}
\<\r r\>_\text{\tiny obst}=10^4 \,\text{g}/ \cm^2\,,
\end{align}
due to surrounding structures, mountains, or other obstacles. Lastly, at angles $\th>\p/2$ the Macro would pass though the Earth and encounter a column density,
\begin{equation}
\<\r r\>_\oplus= 2R_\oplus \r_\oplus \abs{\cos{\th}}\,.
\end{equation}

\ssec{Thermo-acoustic detection of cosmic particles}

Upon its passage through the (identical) NAUTILUS or EXPLORER detectors, a very dense object would deposit energy in a line along its track in the detector. 
Because the Macro velocity is supersonic in the aluminum, 
the energy deposition would be nearly instantaneous.
The resulting near-instantaneous thermal expansion 
would source a pressure wave that would excite the bar's longitudinal vibrational modes. 
As described in e.g.  \cite{Bernard:1983wj, Liu:1988wn, Astone:2013xna}, for a bar of radius $R$ and length $L$, the energy of the lowest mode is
\begin{equation}\label{Vibrationenergy_1}
\Delta E =\f{4}{9\p} \f{\g^2}{\r L v_s^2}\(\f{dE_\text{\tiny X}}{dx}\)^2  G(z_0, l_0, \th_0)\,,
\end{equation}
where $\g$ is the Gr{\"u}neisen parameter, $\r$ is the bar density, $v_s$ is the longitudinal sound speed, and the energy lost by a nuclearite or, more generally, a Macro is
\begin{equation}\label{E-loss_rate}
\abs{\f{dE_\text{\tiny X}}{dx}}=\r \sX \vX^2\,.
\end{equation}
The geometric function that depends on the track through the bar is
\begin{equation}
G(z_0, l_0, \th_0)=\[\f{L}{\p R}\sin{\(\f{\p z_0}{L}\)} \f{\sin{\(\f{\p l_0}{2L} \cos{\th_0}\)}}{\cos{\th_0}}\]^2\,,
\end{equation}
where $z_0$ is the distance of the track midpoint to one end of the bar, $\th_0$ is the angle between the track and the bar axis and $l_0$ is the length of the track. For the bars described in \cite{Astone:2013xna}, $R=0.3\,$m and $L=3$ m; the full geometric acceptance is therefore $19.54$ m$^2\,$sr.


\ssec{Application of nuclearite analysis to Macros}
In Ref. \cite{Astone:2013xna} it is reported that, for the parameters specific to the NAUTILUS and EXPLORER aluminum bars, a vertical impact at the bar center results in an energy deposition (in units of temperature) into the fundamental mode of
\begin{equation}\label{delta_E_Astone}
\Delta E = 10.7~\text{K} \(\f{\vX \,\x(M)}{10^{-3}c}\)^4\,,
\end{equation}
where, in our notation, $\x(M)=(M/1.5\,\text{ng})^{1/3}$ is a function that characterizes the radius of a passing nuclearite as a function of its mass\footnote{For an object of constant density $\sX\propto M^{2/3}$; hence from \eqref{Vibrationenergy_1} and \eqref{E-loss_rate} the $\x(M)^4$ dependence in \eqref{delta_E_Astone} can be understood.}. 

It is straightforward to apply the analysis in \cite{Astone:2013xna} to macro dark matter, considering a nuclearite of mass $1.5$ ng would have a cross section $\s_0\simeq\p \times10^{-16}$ cm$^2$ at the reference density of $3.5\times10^{14}~\gcmc$. The energy deposition by a Macro can then be readily translated by making the replacement
\begin{equation}
\x(M)^2\to \f{\sX}{\s_0}\
\end{equation}
in \eqref{delta_E_Astone}. We will furthermore use $250$ km/s as a reference velocity and reinsert the geometric function, $G$, writing the excitation energy as
\begin{equation}\label{excitation_energy}
\Delta E = 5.2~\text{K} \(\f{\vX}{250 ~\text{km/s}}\)^4 \(\f{\sX}{\s_0}\)^2 G(z_0, l_0, \th_0)\,.
\end{equation}
Given the detection threshold of $2\,$K used in \cite{Astone:2013xna}  and  the Macro velocity upon impact given in \eqref{Macro_velocity}, the fraction of impacts for a given pair of Macro parameters $(\sX,\MX)$ that exceeds this threshold is determined by performing a Monte Carlo simulation over the possible bar impact points and Macro trajectories.

We can, however, give a semianalytic estimate of the constrained region in the $\sX-\MX$ plane by recognizing that at large $\sm$ the only Macros that can make it to the detector with sufficiently high velocity (quantified below) will come from nearly directly overhead. For those impacts $G(z_0, l_0, \th_0)$ is maximized at unity. The requirement for resonant bar detection is then
\begin{equation}
 \(\f{v_0}{250\text{ km/s}}\)^2 e^{-2\f{\sX}{\MX}\<\r r\> }  \(\f{\sX}{\s_0}\) \gtrsim 0.62\,.
\end{equation}
This inequality is saturated at the critical value $\s_\text{\tiny X,c}$, which may be solved implicitly as a function of $\MX$:
\begin{align}
\s_\text{\tiny X,c}=-\f{\MX}{2\<\r r\>}W(x)\,.\label{sX_of_x}
\end{align}
Here
\begin{equation}
x\simeq-1.24\times \<\r r\>\f{\s_0}{\MX } \(\f{v_0}{250\text{ km/s}}\)^{-2} \label{x_def}\,,
\end{equation}
and $W(x)$ is known as the Lambert-W function, defined implicitly by the relation
\begin{equation}
x=W(x)\, \exp{W(x)}\,.
\end{equation}
$W(x)$ has an infinite number of branches; however, the branch $W_{-1}(x)$ delineates the top of the constrained region in the $\sX-\MX$ plane according to \eqref{sX_of_x}. 

We must also ensure that the Macro velocity exceeds the sound velocity of aluminum, $v_s\simeq 2\times10^{-5}\,c$; otherwise the energy loss rate formula \eqref{E-loss_rate} breaks down  \cite{Astone:2013xna}. From \eqref{Macro_velocity} this requires that
\begin{align}
\sX &\lesssim \f{1}{\<\r r\>} \log{\f{v_0}{v_s}} \MX \notag\\
&\lesssim 3.7\times10^{-3}~\cm^2 \(\f{\MX}{1\text{ g}}\) 
\end{align}
where we have used an approximate atmospheric depth\footnote{We have neglected the additional integrated depth the Macro must pass through due to concrete in the roofing of the building in which EXPLORER was housed, as well as the experimental components mounted to the top of the bars \cite{Astone:2008xa}. On the other hand, EXPLORER was itself run at 430 m above sea level at CERN, so we have overestimated the atmospheric depth to it. These two complications (at least) partially offset each other and, in any case, they can be neglected considering the level of accuracy required here.} of $\<\r r\>=10^3$ g/cm$^2$ and $v_0=250$ km/s. For masses larger than roughly $10^{-10}$g this turns out to be more restrictive than using \eqref{sX_of_x}, so it sets the upper edge of the region of constraint in that range. 

Lastly, we consider the Macro flux limitation. At large $\sm$, we cannot use the full geometric acceptance of $\simeq 19.54$ m$^2\,$sr because of significant additional drag that would be experienced by the Macros during their passage through (i) the environment surrounding the detector (e.g. other buildings, mountains, etc.) and (ii) the integrated depth through the Earth. Likewise, Macros with small $\sm$ are more likely to make it to the detector with high enough velocity and energy deposition. Therefore, for the largest $\sm$ we use only 1/4 of the geometric acceptance since we can only trust their ability to make it through the atmosphere. For $\sm\lesssim 10^{-4}\cmsg$ a Macro could easily pass through surrounding obstacles so that using 1/2 of the full acceptance is justified. Finally, for $\sm\lesssim \(\r_\oplus R_\oplus\)^{-1}\simeq3\times10^{-10}\cmsg$ the Macro could pass freely through the Earth and in that case we may use the full acceptance. These approximations ignore the detector efficiency as a function of impact trajectory; however this is taken into account in the Monte Carlo simulation.

\begin{figure}[h!tb]
  \begin{center}
    \includegraphics[scale=.45]{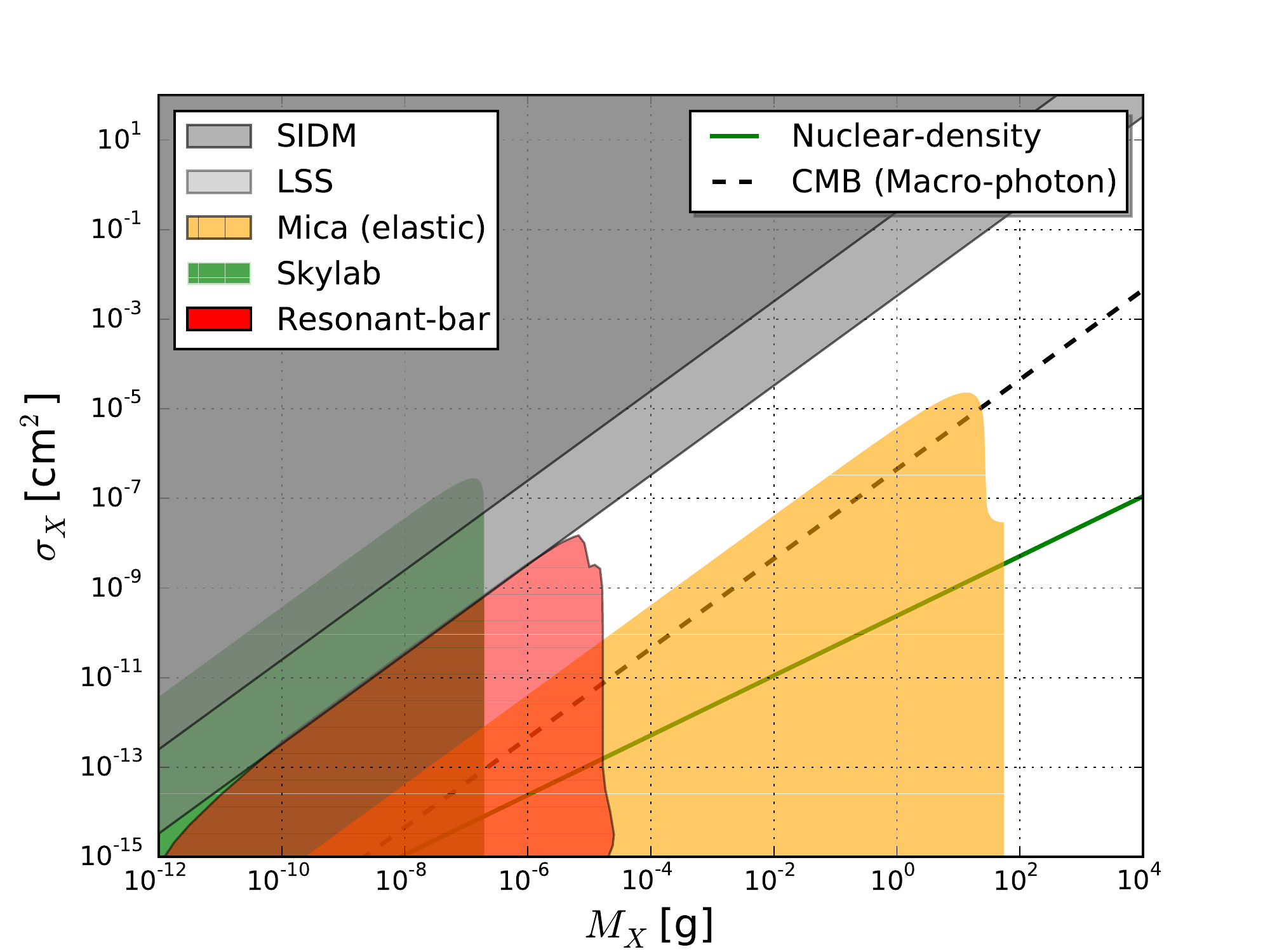}
  \end{center}
\caption{Existing constraints (described in detail in \cite{Jacobs:2014yca}) on macro dark matter that scatters elastically with baryons and our new constraints from resonant bars displayed in red. Note the marginal overlap between the resonant bar and large-scale structure constraints and that slight improvements to the Skylab and mica constraints have been made as compared to \cite{Jacobs:2014yca}; both of these points are explained in the Discussion section. The CMB constraint \cite{Wilkinson:2013kia, Jacobs:2014yca}, applicable if Macros couple to photons, rules out Macros above the dashed line.}
\label{gwave_plot}
\end{figure}

Given an isotropic dark matter flux of
\begin{align}
\f{1}{4\p}\nX v_0 \simeq& 1.2\times10^{-9}\,\text{m}^{-2}\text{sr}^{-1}\text{day}^{-1}\notag\\
~~~&\times
\(\f{1\text{~g}}{\MX}\)\(\f{\rX}{0.4\,\GeV/\text{cm}^3}\)\(\f{v_0}{250\text{~km/s}}\)
\end{align}
and the combined 3,921 live-time days of NAUTILUS and EXPLORER, the maximum number of events expected is
\begin{equation}
N\simeq 10^{-4} \(\f{1\text{~g}}{\MX}\)
\end{equation}
for a local dark matter density, $\rX=0.4\,\GeV/\text{cm}^3$, and $v_0=250\text{~km/s}$. A Macro impact would be a random (Poisson) process; therefore a null detection of Macro passages for which  $N>3$ events are expected will rule them out at greater than 95\% confidence. It follows that, accounting for the usable fraction of the geometric acceptance mentioned above, the following macro dark matter candidates are ruled out by this analysis:
\begin{equation}
\MX< 
	\begin{cases}
		3 \times10^{-5} 	\text{g}, 
		& \text{if }  \frac{\sX}{\MX}\lesssim10^{-9}\frac{\cm^2}{\text{g}};\\
		2\times10^{-5} 	\text{g}, 
		& \text{if } 10^{-9}\frac{\cm^2}{\text{g}}\lesssim\,\frac{\sX}{\MX}\lesssim10^{-4}\frac{\cm^2}{\text{g}};\\
		10^{-5} 		\text{g}, 
		& \text{if } 10^{-4}\frac{\cm^2}{\text{g}}\lesssim\,\frac{\sX}{\MX}\lesssim10^{-3}\frac{\cm^2}{\text{g}}.
	\end{cases}
\end{equation}
Note that at the smallest $\sm$ our bound differs slightly from the one found in \cite{Astone:2013xna} because of both our choice of $v_0=250\text{~km/s}$ and 95\% confidence requirement. 
Deviations from these estimates are expected at low enough $\sX$ because of the decreasing excitation energy as a function of $\sX$, as seen in \eqref{excitation_energy}. The results of our more accurate Monte Carlo analysis are illustrated in Fig. \ref{gwave_plot}, where, however, we have used the fixed value of $v_0=250$ km/s.

\sec{Discussion}

Here we have been able to extend existing constraints on nuclearites \cite{Astone:2013xna}, based on resonant bar gravitational wave detectors, to place new constraints on macro dark matter coupling to baryons. If Macros also couple to photons with their geometric cross section then the cosmic microwave background (CMB) constraints indicating  $\sm\lesssim4.5\times 10^{-7}\cmsg$ (dashed line in Fig. \ref{gwave_plot}), when taken in combination with the mica constraints, completely overlap with these constraints. Nevertheless, our results are relevant since they directly constrain Macro-baryon coupling which could, in principle, be different from Macro-photon coupling. Also, the resonant bar detectors have acted as a local probe that directly constrains Macro properties in the solar neighborhood without relying on the (albeit reasonable) assumption that the local dark matter is of similar form and composition to dark matter in different parts and epochs of the Universe.

Here the velocity distribution of Macros was taken to be isotropic; in reality the Earth (with the Solar System) moves through the Galaxy at approximately 200-250 km/s and this should result in an anisotropic distribution observed on the Earth. Also, the Macro velocity distribution presumably has a tail from which it is common to draw an initial Macro velocity, $v_0$ significantly larger than $250$ km/s. This means that the constraints inferred here are conservative and the region of constraint should extend upwards to larger $\sX$, i.e. the constraint is stronger than $\sm\lesssim 3.7\times 10^{-3} \cmsg$. Therefore, greater overlap is expected with those determined from large-scale structure, which are $\sm\lesssim 3.3\times 10^{-3} \cmsg$ \cite{Dvorkin:2013cea}. On the other hand, the range of $\sm$ constrained here is only log-sensitive to $v_0$, so we do not expect this to significantly enhance our results.

Here we have also made modest improvements to the constraints inferred from both Skylab \cite{shirk1978charge} and ancient mica \cite{Price:1988ge} compared to those presented in \cite{Jacobs:2014yca}. For both detectors we have more carefully calculated their acceptance, taking into account the column density encountered by a Macro as a function of its impact angle--this resulted in more rounded corners of the constrained regions. For the mica samples studied in \cite{Price:1988ge} the orientation during their $\sim500\,$Myr exposure time is unknown; we therefore computed the minimum region that is ruled out for a flat mica sample oriented either parallel or perpendicular to the Earth's surface and have presented those constraints in Fig. \ref{gwave_plot}. \

The new constraints presented here on macro dark matter from resonant bar detectors provide overlap with both the Skylab and mica constraints, and also fill in part of the gap between the large-scale structure and mica bounds on the coupling of Macros to baryons. In future studies, however, a dedicated experiment of this type is far from ideal due to its relatively limited exposure, which is proportional to the detector surface area and experiment lifetime.
\\\\
\emph{Acknowledgements.}--The authors thank Tony Walters and Vincent Bouillot for discussions. One of the authors (D.M.J.) would like to acknowledge support from the Claude Leon Foundation.  G.D.S. acknowledges support by Department of Energy Grant No. DOE-SC0009946. This work is based on research supported in part by the National Research Foundation of South Africa (Grants No. 91552 and No. 81134). D.M.J. thanks CWRU, A.W. thanks NITheP and G.D.S. thanks UCT for hospitality during stages of this research.
 
\newpage

\bibliographystyle{h-physrev}
\bibliography{Macro_resonant_bib}

\end{document}